\documentclass[prd,amsmath,amssymb,superscriptaddress,preprintnumbers,twocolumn,10pt]{revtex4-1}

\pdfoutput=1

\usepackage{graphicx}
\usepackage{dcolumn}
\usepackage{bm}
\usepackage{amssymb}
\usepackage{latexsym}
\usepackage{booktabs}
\usepackage{amsmath}
\usepackage{multirow}
\usepackage{url}

\usepackage{float}
\usepackage[colorlinks=true, linkcolor=red, citecolor=blue]{hyperref}

\usepackage[normalem]{ulem}
\usepackage{color}
\usepackage{array}
\usepackage{enumerate}
\usepackage{setspace} % spacing

\begin{document}

%\setstretch{1.2} % line spacing

\title{Joint constraints on cosmological parameters using future multi-band gravitational wave standard siren observations}

\author{Shang-Jie Jin}
\affiliation{Key Laboratory of Cosmology and Astrophysics (Liaoning Province) \& Department of Physics, College of Sciences, Northeastern University, Shenyang 110819, China}
\author{Shuang-Shuang Xing}
\affiliation{Key Laboratory of Cosmology and Astrophysics (Liaoning Province) \& Department of Physics, College of Sciences, Northeastern University, Shenyang 110819, China}
\author{Yue Shao}
\affiliation{Key Laboratory of Cosmology and Astrophysics (Liaoning Province) \& Department of Physics, College of Sciences, Northeastern University, Shenyang 110819, China}
\author{Jing-Fei Zhang}
\affiliation{Key Laboratory of Cosmology and Astrophysics (Liaoning Province) \& Department of Physics, College of Sciences, Northeastern University, Shenyang 110819, China}
\author{Xin Zhang}\thanks{Corresponding author.\\zhangxin@mail.neu.edu.cn}
%\email{zhangxin@mail.neu.edu.cn}
\affiliation{Key Laboratory of Cosmology and Astrophysics (Liaoning Province) \& Department of Physics, College of Sciences, Northeastern University, Shenyang 110819, China}
\affiliation{Key Laboratory of Data Analytics and Optimization for Smart Industry (Ministry of Education),
Northeastern University, Shenyang 110819, China}
\affiliation{National Frontiers Science Center for Industrial Intelligence and Systems Optimization,
Northeastern University, Shenyang 110819, China}
%\date{\today}

\begin{abstract}
Gravitational waves (GWs) from the compact binary coalescences can be used as standard sirens to explore the cosmic expansion history. In the next decades, it is anticipated that we could obtain the multi-band GW standard siren data (from nanohertz to a few hundred hertz), which are expected to play an important role in cosmological parameter estimation. In this work, we give for the first time the joint constraints on cosmological parameters using the future multi-band GW standard siren observations. We simulate the multi-band GW standard sirens based on the SKA-era pulsar timing array (PTA), the Taiji observatory, and the Cosmic Explorer (CE) to perform cosmological analysis. In the $\Lambda$CDM model, we find that the joint PTA+Taiji+CE data could provide a tight constraint on the Hubble constant with a $0.5\%$ precision. Moreover, PTA+Taiji+CE could break the cosmological parameter degeneracies generated by CMB, especially in the dynamical dark energy models. When combining the PTA+Taiji+CE data with the CMB data, the constraint precisions of $\Omega_{\rm m}$ and $H_0$ are $1.0\%$ and $0.3\%$, meeting the standard of precision cosmology. The joint CMB+PTA+Taiji+CE data give $\sigma(w)=0.028$ in the $w$CDM model and $\sigma(w_0)=0.11$ and $\sigma(w_a)=0.32$ in the $w_0w_a$CDM model, which are comparable with or close to the latest constraint results by CMB+BAO+SN. In conclusion, it is worth expecting to use the future multi-band GW observations to explore the nature of dark energy and measure the Hubble constant.

\end{abstract}

\maketitle

\section{Introduction}\label{sec:intro}

The precise measurements of the cosmic microwave background (CMB) anisotropies initiated the era of precision cosmology \cite{WMAP:2003elm,WMAP:2003ivt}. Nevertheless, with the improvement of measurement precisions of cosmological parameters, some tensions between the early- and late-universe observations arised.
In particular, the values of the Hubble constant inferred from the $Planck$ CMB observation (based on the $\Lambda$CDM model) \cite{Planck:2018vyg} and determined by the distance-ladder measurement (model-independent) \cite{Riess:2021jrx} are shown to be in more than $5\sigma$ tension \cite{Riess:2021jrx}, which is now commonly believed as a severe crisis for cosmology \cite{Riess:2019qba,Verde:2019ivm}. The Hubble tension is widely discussed in the literature \cite{Riess:2019qba,Verde:2019ivm,Guo:2018ans,Perivolaropoulos:2021jda,Gao:2021xnk,DiValentino:2021izs,Abdalla:2022yfr,Cai:2021wgv,Yang:2018euj,DiValentino:2020zio,DiValentino:2019jae,DiValentino:2019ffd,Liu:2019awo,Zhang:2019cww,Ding:2019mmw,Li:2020tds,Wang:2021kxc,Vagnozzi:2021tjv,Vagnozzi:2021gjh,Vagnozzi:2019ezj,Guo:2019dui,Vagnozzi:2019ezj,Feng:2019jqa,Lin:2020jcb,2022arXiv221213146G,2022arXiv221213433Z,Liu:2021jnw}. So far, there is no consensus on a valid extended cosmological model that can truly solve the Hubble tension. Therefore, some cosmological probes that can independently measure the Hubble constant need to be greatly developed. The gravitational wave (GW) standard siren method is one of the most promising options.

Different from the traditional electromagnetic (EM) observations, GW observations open a new window into exploring the expansion history of the universe. The GW waveform encodes the information of the luminosity distance, which is called a standard siren \cite{Schutz:1986gp,Holz:2005df}. Applying GW standard sirens in cosmology has recently been widely discussed in the literature \cite{Holz:2005df,Dalal:2006qt,Nissanke:2009kt,Cutler:2009qv,Camera:2013xfa,Vitale:2018wlg,Bian:2021ini,Cai:2016sby,Cai:2017aea,Cai:2017plb,Zhang:2019ylr,Chen:2020dyt,Gray:2019ksv,Zhao:2010sz,Zhao:2018gwk,Du:2018tia,Cai:2018rzd,Yang:2019bpr,Yang:2019vni,Bachega:2019fki,Chang:2019xcb,Zhang:2019loq,Mukherjee:2019qmm,He:2019dhl,Zhao:2019gyk,Wang:2021srv,Qi:2021iic,Jin:2021pcv,Zhu:2021bpp,deSouza:2021xtg,2022arXiv220100607W,2022arXiv220209726W,Jin:2022tdf,2022arXiv221110087H,2022arXiv220813999C,Wang:2022rvf,2022arXiv221213183D,2022arXiv220303956C,2022arXiv221200531S,Cao:2021zpf,Leandro:2021qlc,Fu:2021huc,Ye:2021klk,Chen:2020zoq,Mitra:2020vzq,Hogg:2020ktc,Nunes:2020rmr,Borhanian:2020vyr,2022arXiv220211882J,Jin:2020hmc,Ghosh:2022muc,Yu:2020vyy,Wang:2020xwn}. If the redshift information of the GW source could be obtained by identifying the EM counterparts (we usually refer to this kind of GW standard sirens as bright sirens), the distance-redshift relation could be established for cosmological parameter estimations. While for the GW events without EM counterparts, the statistical analysis of the GW event associated with the galaxy catalog can also be applied in obtaining the redshift information (we usually refer to this kind of GW standard sirens as dark sirens).

In fact, the frequency ranges of GW standard sirens are wide (from nanohertz to a few hundred hertz), corresponding to different GW sources. Aiming at detecting GWs in different frequency bands, the pulsar timing arrays (PTAs), the space-based GW detectors, and the ground-based GW detectors are proposed.

The nanohertz GWs emitted by the supermassive black hole binaries (SMBHBs) could be detected by PTA, a natural Galactic-scale detector of millisecond pulsars (MSPs). Although it is difficult to detect GWs from individual SMBHBs by the current PTA projects, e.g., the European Pulsar Timing Array \cite{Kramer:2013kea}, the North American Nanohertz Observatory for Gravitational Waves \cite{McLaughlin:2013ira}, and the Parkes Pulsar Timing Array (Australia) \cite{Hobbs:2013aka}, it is expected that the individual SMBHBs could be detected by the SKA-era PTAs \cite{Smits:2008cf}. Yan \emph{et al.} \cite{2020ApJ...889...79Y} proposed that the currently available SMBHB candidates with known redshifts could be detected by the future SKA-era PTAs, allowing SMBHBs to be treated as standard sirens to explore the cosmic expansion history. Wang \emph{et al.} \cite{2022arXiv220100607W} forecasted the cosmological parameter estimation with the bright sirens and dark sirens of individual SMBHBs with the SKA-era PTAs.

The space-based GW detectors are proposed to detect GWs emitted from the massive black hole binaries (MBHBs) in the millihertz frequency band, e.g., Taiji \cite{Wu:2018clg,Ruan:2018tsw,Hu:2017mde}, TianQin \cite{Luo:2020bls,Milyukov:2020kyg,TianQin:2020hid}, and the Laser Interferometer Space Antenna \cite{LISA:2017pwj,LISACosmologyWorkingGroup:2022jok}. The space-based GW detectors could detect high-redshift GW events (up to $z\simeq 15-20$), which are expected to provide high-redshift GW standard siren data. Some works show that the EM signals could be emitted in the process of MBHB mergers in both the radio and optical bands \cite{Palenzuela:2010nf,OShaughnessy:2011nwl,Moesta:2011bn,Kaplan:2011mz,Shi:2011us,Blandford:1977ds,Meier:2000wk,Dotti:2011um}. The applications of these bright sirens in cosmological parameter estimation have been forecasted in the literature \cite{LISACosmologyWorkingGroup:2022jok,Zhao:2019gyk,Wang:2019tto,Wang:2021srv,Zhu:2021bpp,Zhu:2021aat,Mangiagli:2022elu,Tamanini:2016uin,Caprini:2016qxs}.

The ground-based GW detectors could observe stellar-mass binaries in the frequency band of a few hundred hertz.
The only multi-messenger observation event GW170817 from a binary neutron star (BNS) merger gave the first measurement of the Hubble constant using the standard siren method with a $14\%$ precision \cite{LIGOScientific:2017adf}. The measurement precision of the Hubble constant could reach $2\%$ using 50 similar GW standard sirens \cite{Chen:2017rfc}, showing the potential of standard siren method in cosmological parameter estimation.
{While for the dark siren method, the latest constraint precision of the Hubble constant from the LIGO-Virgo-KAGRA observation is 19\% \cite{LIGOScientific:2021aug} (recent related works can refer to, e.g., Refs.~\cite{DES:2019ccw,DES:2020nay,LIGOScientific:2019zcs}).}
In the next decades, the third-generation (3G) ground-based GW detectors, the Cosmic Explorer (CE) in the U.S. \cite{LIGOScientific:2016wof} and the Einstein Telescope (ET) in Europe \cite{Punturo:2010zz}, will observe a large number of GW events in a wide range of redshift because the sensitivities of them are one order of magnitude improved over the current detectors \cite{Evans:2021gyd}.

In the next decades, it is expected that we could obtain the multi-band GW standard siren data. Owing to the fact that the numbers of detectable GWs and signal-to-noise ratios (SNRs) in different frequency bands are different, the joint future multi-band GW standard siren observations are expected to play an important role in cosmological parameter estimation.

In this work, the first question to be answered is what precision the cosmological parameters could be measured to by the joint constraints of future multi-band GW standard siren observations. The second question we wish to answer is what role the multi-band GW standard sirens could play in breaking cosmological parameter degeneracies generated by the EM observations. Note that, in this work, we only focus on the GW bright standard siren observations. We will consider the future bright siren observations from the SKA-era PTAs, the space-based GW detectors, and the 3G ground-based GW detectors, which are in different frequency bands, and constrain the cosmological parameters relevant to dark energy and the Hubble constant issues using the mock data of joint multi-band GW standard sirens.

The paper is organized as follows. In Section~\ref{simulation:PTA}, we introduce the method of simulating GW standard sirens from the SKA-era PTA. In Section~\ref{simulation:Taiji}, we introduce the method of simulating GW standard sirens from Taiji. In Section~\ref{simulation:CE}, we introduce the method of simulating GW standard sirens from CE.
In Section~\ref{results}, we give the constraint results and make some relevant discussions. The conclusion is given in Section~\ref{Conclusion}. {We adopt the $\Lambda$CDM model as the fiducial model to generate the simulated GW standard siren data, with the cosmological parameters set to the constraint results obtained from $Planck$ 2018 TT,TE,EE+lowE \cite{Planck:2018vyg}.}

\section{Method}

\subsection{Simulation of GW standard sirens from SKA-era PTAs}\label{simulation:PTA}

GW signals are detected in the timing residuals of MSPs by removing model-predicted times of arrival (ToAs) from the observational ToA data. The time residuals induced by a single GW source measured at time $t$ on the Earth can be written as \cite{2020ApJ...889...79Y}
\begin{align}
s(t,\hat{\Omega}_{\rm s},\hat{\Omega}_{\rm p})=F_{+}(\hat{\Omega}_{\rm s},\hat{\Omega}_{\rm p})\Delta A_+(t)+F_{\times}(\hat{\Omega}_{\rm s},\hat{\Omega}_{\rm p})\Delta A_{\times}(t),
\end{align}
where $F_{+, \times}(\hat{\Omega}_{\rm s},\hat{\Omega}_{\rm p})$ are the antenna response functions \cite{Wahlquist:1987rx}, $\hat{\Omega}_{\rm s}$ and $\hat{\Omega}_{\rm p}$ are the unit vectors pointing from the GW source ($\alpha_{\rm s}$, $\beta_{\rm s}$) and pulsar to the observer ($\alpha_{\rm p}$, $\beta_{\rm p}$), respectively. $\Delta A_{+,\times}(t)=A_{+,\times}(t)-A_{+,\times}(t_{\rm p})$ is the difference between the earth term $A_{+,\times}(t)$ and the pulsar term $A_{+,\times}(t_{\rm p})$, with $t_{\rm p}$ the time at which GW passes the MSP \cite{Ellis:2012zv}. The forms of $A_{+,\times}(t)$ are related to the GW strain. Assuming SMBHBs inspiral in circular orbits, the GW strain $h(t)$ can be written as
\begin{align}
h(t)=&2\frac{(G\mathcal M_{\rm c})^{5/3}}{c^{4}}\frac{[\pi f(t)]^{2/3}}{d_{\rm L}},\\
f(t)=&[f_{0}^{-8/3}-\frac{256}{5}\pi^{8/3}(\frac{G\mathcal{M}_{\rm c}}{c^{3}})^{5/3}t]^{-3/8},
\end{align}
where $\mathcal{M}_{\rm c}$ is the observed chirp mass, $M=m_1+m_2$ is the total mass of a binary system with the component masses $m_1$ and $m_2$, $\eta=m_1 m_2/(m_1+m_2)^2$ is the symmetric mass ratio, $f_0=2f_{\rm orb}$ is the GW frequency at the time of our first observation, $f_{\rm orb}=(2\pi T)^{-1}$ is the orbit frequency, and $T$ is the orbital periods of SMBHB candidates taken from Refs.~\cite{Valtonen:2008tx,Graham:2015gma,Graham:2015tba,Charisi:2016fqw,Yan:2015mya,Li:2016hcm,Zheng:2015dij,Li:2017eqf}. Here we calculate $f_{0}$ using the orbital periods of the 154 SMBHB candidates.

The SNR of the GW signal detected by a PTA is written as \cite{2020ApJ...889...79Y}
\begin{align}
\rho^2=\sum_{i=1}^{N_{\rm p}}\sum_{n=1}^{N}\left[\frac{s_i(t_n)}{\sigma_{t,i}}\right]^2,
\end{align}
where $N_{\rm p}$ is the number of MSPs, $N$ is the total number of data points of each MSP, $s_{i}(t_{n})$ is the timing residual of $i$-th MSP at time $t_{n}$, and $\sigma_{t,i}$ is the root mean square (rms) timing residual of the $i$-th MSP. Here we set the threshold of SNR to be 8.

We use the Fisher information matrix to estimate measurement errors of $d_{\rm L}$. For a PTA containing $N_{\rm p}$ independent MSPs, the Fisher matrix is expressed as \cite{2020ApJ...889...79Y}
\begin{align}
{F}_{ab}=\sum_{i=1}^{N_{\rm p}}\sum_{n=1}^{N}\frac{\partial{s_i(t_n)}}{\sigma_{t,i}\partial{\theta_a}}\frac{\partial{s_i(t_n)}}{\sigma_{t,i}\partial{\theta_b}},
\end{align}
where $\bm{\theta}$ denotes the free parameters to be estimated. Here, the Fisher matrix includes nine parameters, including eight GW source parameters ($d_{\rm L}$, $\mathcal{M}_{\rm c}$, $\alpha_{\rm s}$, $\beta_{\rm s}$, $\iota$, $\psi$, $\phi_{0}$, $f_{0}$) and the pulsar distance $d_{\rm p}$. The error of the parameter $\theta_a$ is calculated by $\Delta\theta_a=\sqrt{(F^{-1})_{aa}}$, i.e., $\sigma_{d_{\rm L}}^{\rm inst}=\Delta d_{\rm L}=\Delta\theta_1$.

The measurement of $d_{\rm L}$ is also affected by the weak lensing and we adopt the form \cite{Tamanini:2016zlh,Speri:2020hwc,Hirata:2010ba}
\begin{align}
\sigma_{d_{\rm L}}^{\rm lens}(z)=&\left[1-\frac{0.3}{\pi / 2} \arctan \left(z / 0.073\right)\right]\nonumber\\
&\times d_{\rm L}(z)\times 0.066\bigg[\frac{1-(1+z)^{-0.25}}{0.25}\bigg]^{1.8}.\label{lens}
\end{align}
The total error of $d_{\rm L}$ can be written as $\sigma_{d_{\rm L}}=\sqrt{(\sigma_{d_{\rm L}}^{\rm inst})^2+(\sigma_{d_{\rm L}}^{\rm lens})^2}$.

We analyze a catalog of 154 currently available SMBHB candidates for this work. Among them, 149 are obtained through periodic variations in their light curves, as described in previous studies \cite{Graham:2015gma,Graham:2015tba,Charisi:2016fqw}. The remaining candidates include Mrk 231 from Ref.~\cite{Yan:2015mya}, NGC 5548 from Ref.~\cite{Li:2016hcm}, OJ 287 from Ref.~\cite{Valtonen:2008tx}, SDSS J0159+0105 from Ref.~\cite{Zheng:2015dij}, and Ark 120 from Ref.~\cite{Li:2017eqf}.
We obtain the right ascension, declination, redshift, and total mass information for the SMBHB candidates and fix other parameters. Previous work has shown that the polarization angle $\psi$ and initial phase $\phi_0$ have no significant effects on GW analysis \cite{2020ApJ...889...79Y}. Therefore, we follow Refs.~\cite{2020ApJ...889...79Y,2022arXiv220100607W} and assume $\psi = 0^{\circ}$ and $\phi_0 = 0^{\circ}$ in our simulation.
For the inclination angle $\iota$, we assume that all the GW events have an edge-on inclination angle, i.e., $\iota=90^{\circ}$. While this assumption may not always hold true, we adopt it as a conservative analysis. In addition, we wish to note that we make an optimistic assumption that the mass ratios of the SMBHB candidates are assumed to be $q=1$, which is also adopted in Refs.~\cite{2020ApJ...889...79Y,2022arXiv220100607W}. In fact, the mass ratio can impact the strength of the GW signal, with smaller values of $q$ leading to weaker GW signals, larger measurement errors of $d_{\rm L}$, and worse constraint results, as discussed in previous literature (e.g., Ref.~\cite{2020ApJ...889...79Y}).

Here we note that the ability of PTA GW observations is affected by many factors, for example, the number of MSPs $N_{\rm p}$ and rms of time residual $\sigma_t$. It is found that about 100 high-quality MSPs are sufficient for the detection of nanohertz GWs from individual SMBHBs \cite{2022arXiv220100607W}. {The current PTAs usually contain dozens of MSPs. In the future, we can expect SKA and FAST to observe hundreds of MSPs.} In the present work, we simulate 200 pulsars to detect nanohertz GWs. For the rms of timing residual, it consists of white noise and red noise.
Recent analysis shows that the total white noises of pulsars {could approach} 10--50 ns \cite{Porayko:2018sfa} in the SKA era, thus here we consider $\sigma_{t}$ to be $\sigma_{t}=20$ ns.
{SMBHB candidates usually emit GWs in the frequency range of $10^{-7}$--$10^{-8}$ Hz. In this frequency,} the red noise can be attenuated to a low noise level, and it does not affect the single GW detection, so it can be ignored.
{In addition, the stochastic gravitational wave background (SGWB) will also affect the detection of SMBHBs. However, recent studies have shown that SGWB is likely to be detected in about ten years. The SGWB can be regarded as red noise, which has slight impact in the frequency of $10^{-7}$--$10^{-8}$ Hz.}
Following Ref.~\cite{2022arXiv220100607W}, we assume that the ToA data are obtained by monitoring the MSPs with the typical cadence of two weeks and the observation time is 10 years.
{Based on the simulation method introduced above, we simulate 35 bright sirens for the 10-year observation of SKA-era PTAs, which are shown in Fig.~\ref{Fig2}. Note that the number of detected SMBHBs in the SKA era is expected to be much larger than 154. A full analysis of the expected detection number of SMBHBs in the SKA era is left to a future work.}

\subsection{Simulation of GW standard sirens from Taiji}
\label{simulation:Taiji}
In this section, we focus on the GW signal from the inspiral of a non-spinning MBHB. The frequency domain GW waveform is written as
\begin{eqnarray}
\tilde{h}(f)=-\left(\frac{5\pi}{24}\right)^{1/2}\mathcal{M}^{5/6}_{\rm c}\left[\frac{(\pi f)^{-7/6}}{D_{{\rm eff}}}\right] e^{-i\Psi}.
\end{eqnarray}
The effective luminosity distance, denoted as $D_{\rm eff}$, is given by the formula $D_{{\rm eff}}=d_{\rm L} [F^{2}_{+}(\frac{1+{\rm cos}^2 \iota}{2})^2+F^{2}_{\times} {\rm cos}^2 \iota]^{-1/2}$. Here, $d_{\rm L}$ is the luminosity distance, and $F_{+,\times}$ are the antenna response functions that depend on the location of the GW source ($\theta$, $\phi$) and the polarization angle $\psi$. The detailed expressions for $F_{+,\times}$ can be found in Ref.~\cite{2019arXiv190907104R}. The GW phase $\Psi$ is written to the second Post-Newtonian order and is related to the coalescence time $t_{\rm c}$ and the coalescence phase $\psi_{\rm c}$. The specific form of $\Psi$ can be found in Ref.~\cite{2019arXiv190907104R}. To describe the GW signal in Fourier space, the observation time $t$ is replaced with $t(f)=t_{\rm c}-\frac{5}{256}\mathcal{M}^{-5/3}_{\rm c}(\pi f)^{-8/3}$ \cite{Buonanno:2009zt,Krolak:1995md}.

The combined SNR for the detection network of $N$ independent interferometers is given by
\begin{equation}
\rho=\sqrt{\sum_{i=1}^{N}(\rho_i)^2},\label{snr}
\end{equation}
where $\rho_i=\sqrt{\langle\tilde{h_i},\tilde{h_i}\rangle}$. The inner product is defined as
\begin{equation}
\langle a,b\rangle=2\int_{f_{\rm lower}}^{f_{\rm upper}}\frac{a(f)b^{*}(f)+a^{*}(f)b(f)}{S_{\rm n}(f)}df,
\end{equation}
where $f_{\rm lower}=10^{-4}$ Hz and $f_{\rm upper}=c^{3}/6\sqrt{6}\pi GM_{\rm obs}$ with $M_{\rm obs}=(m_{1}+m_{2})(1+z)$ \cite{Feng:2019wgq}.
Taiji's PSD is taken from Ref.~\cite{Ruan:2018tsw}. We set the threshold of SNR to be 8 in the simulation.

For a network with $N$ independent interferometers, the Fisher matrix can be written as
\begin{equation}
F_{ab}=\sum_{i=1}^{N}\left\langle\frac{\partial\tilde{h}_{i}}{\partial \theta_a},  \frac{\partial\tilde{h}_{i}}{\partial \theta_b}\right\rangle,
\end{equation}
where $\bm{\theta}$ denotes nine GW source parameters ($d_{\rm L}$, $\mathcal{M}_{\rm c}$, $\eta$, $t_{\rm c}$, $\psi_{\rm c}$, $\iota$, $\theta$, $\phi$, $\psi$). The instrumental error of the luminosity distance is $\sigma_{d_{\rm L}}^{\rm inst}=\Delta d_{\rm L}=\sqrt{(F^{-1})_{11}}$.
We use Eq.~(\ref{lens}) to calculate the weak-lensing error.
The error caused by the peculiar velocity of the GW source is given by \cite{Kocsis:2005vv}
\begin{equation}
\sigma_{d_{\rm L}}^{\rm pv}(z)=d_{\rm L}(z)\times \big[ 1+ \frac{c(1+z)^2}{H(z)d_{\rm L}(z)}\big]\frac{\sqrt{\langle v^2\rangle}}{c},
\end{equation}
where $H(z)$ is the Hubble parameter and $\sqrt{\langle v^{2}\rangle}$ is the peculiar velocity of the GW source. In this work, we set $\sqrt{\langle v^{2}\rangle}=500$ km $\cdot$ s$^{-1}$, in agreement with average values observed in galaxy catalogs \cite{Speri:2020hwc}.
{In addition, we make the assumption that redshift measurements at $z<2$ are determined spectroscopically with negligible errors (see, e.g., Refs.~\cite{Dawson:2015wdb,Amendola:2016saw}). However, for the GW event with $z>2$ associated with photometric measurements, the redshift error is estimated as $\Delta z\approx 0.03(1+z)$ \cite{Dahlen:2013fea,Ilbert:2013bf}. Following Refs.~\cite{Tamanini:2016zlh,Speri:2020hwc,Zhao:2019gyk,Wang:2019tto,Wang:2021srv}, we propagate the redshift error to the distance error by assuming our fiducial cosmology, i.e., $\sigma^{\rm reds}_{d_{\rm L}}=\frac{\partial d_{\rm L} }{\partial z}\Delta z$. In fact, as shown in Fig.~\ref{Fig1}, for the GW events at $z>2$, the error from lensing is dominant. Therefore, the treatment has little effect on cosmological parameter estimation.}
The total error of $d_{\rm L}$ can be written as $\sigma_{d_{\rm L}}=\sqrt{(\sigma_{d_{\rm L}}^{\rm inst})^2+(\sigma_{d_{\rm L}}^{\rm lens})^2+(\sigma_{d_{\rm L}}^{\rm pv})^2+(\sigma_{d_{\rm L}}^{\rm reds})^2}$.

Owing to the fact that the origin of MBHs is currently unclear, there is uncertainty in predicting the event rate of MBHBs. Based on a semianalytical galaxy formation model, three population models of MBHBs, the pop III, Q3d, and Q3nod models are proposed. The three models have different mechanisms of seeding and delay. In fact, previous works \cite{Tamanini:2016zlh, Wang:2021srv} have shown that the three population models can lead to different constraints on cosmological parameters due to the difference in the number of standard sirens based on the three models. The pop III model typically offers intermediate constraints among the three cases. Therefore, in the present work, we generate simulated standard siren data based on the pop III model.
Some works show that the EM signals could be emitted in the process of MBHB mergers in both the radio and optical bands \cite{Palenzuela:2010nf,OShaughnessy:2011nwl,Moesta:2011bn,Kaplan:2011mz,Shi:2011us,Blandford:1977ds,Meier:2000wk,Dotti:2011um}. Recent works predicted the number of GW detected by space-based GW observatory whose EM counterparts could also be detected by SKA, ELT, and LSST \cite{Tamanini:2016zlh,Wang:2021srv,Yang:2021qge,Mangiagli:2022niy,Mangiagli:2022elu}. {Following Refs.~\cite{Tamanini:2016zlh,Wang:2021srv,Yang:2021qge,Mangiagli:2022niy,Mangiagli:2022elu}, we select the GW events with $\rm SNR>8$ and the sky localization error of $\Delta \Omega<10~\rm deg^2$ (corresponding to the field of view of LSST) as potential candidates for EM counterpart detections. Previous work has shown that the EM counterparts detected by LSST are also detectable for SKA+ELT \cite{Tamanini:2016zlh}. Therefore, in the present work, we only consider the case of SKA+ELT. In this case, EM counterparts may first be detected by SKA in the radio band and the host galaxies are then identified through localization \cite{Tamanini:2016zlh}. Then, the redshifts are determined spectroscopically or photometrically by the optical telescope ELT. We calculate the radio and optical luminosities of an EM counterpart, and it can be detected if its luminosities meet the thresholds of SKA and ELT, as discussed in, e.g., Refs.~\cite{Tamanini:2016zlh,Wang:2021srv,Mangiagli:2022niy}.}

For each simulated standard siren event, the sky location ($\theta$, $\phi$), the binary inclination $\iota$, the coalescence phase $\psi_{\rm c}$, and the polarization angle $\psi$ are evenly sampled in the ranges of $\cos\theta$ $\in$ $[-1, 1]$, $\phi$ $\in$ $[0, 360^{\circ}]$, $\cos\iota$ $\in$ $[-1, 1]$, $\psi_{\rm c}$ $\in$ [0, $ 360^\circ$], $\psi$ $\in$ [0, $360^\circ$], respectively. In this work, we assume $t_{\rm c}=0$ for simplicity. For the redshift and mass distributions of MBHBs, we use the numerical fitting formulas in Ref.~\cite{Wang:2021srv} to fit the curves shown in Figure~3 of Ref.~\cite{Klein:2015hvg}.
{According to the simulation method introduced above, we simulate 28 bright sirens for the 5-year observation of Taiji based on the pop III model, which are shown in Fig.~\ref{Fig2}. The number is also basically consistent with that given in Ref.~\cite{Wang:2021srv}.}

\subsection{Simulation of GW standard sirens from CE}
\label{simulation:CE}

In this work, we consider that all the GW standard sirens that can be detected by CE are the BNS mergers. The redshift distribution of BNS mergers adopts the form in Refs.~\cite{Zhao:2010sz,Cai:2016sby}. we adopt the GW waveform of the inspiralling non-spinning BNS system \cite{Zhao:2010sz}.
We use Eq.~(\ref{snr}) calculate the SNR of each GW event. For CE, $f_{\rm lower}=1$ Hz is the lower cutoff frequency and $f_{\rm upper}=2/(6^{3/2}2\pi M_{\rm obs})$ is the frequency at the last stable orbit with $M_{\rm obs}=(m_1+m_2)(1+z)$. We adopt the PSD of 40 km-arm-length CE \cite{CE-web}. Here we set the threshold of SNR to 8 in our simulation.
In this work, we consider three measurement errors of $d_{\rm L}$, including the instrumental error $\sigma^{\rm inst}_{d_{\rm L}}$, the weak-lensing error $\sigma^{\rm lens}_{d_{\rm L}}$, and the peculiar velocity error $\sigma^{\rm pv}_{d_{\rm L}}$.
%The total error of $d_{\rm L}$ is $\sigma_{d_{\rm L}}=\sqrt{(\sigma_{d_{\rm L}}^{\rm inst})^2+(\sigma_{d_{\rm L}}^{\rm lens})^2+(\sigma_{d_{\rm L}}^{\rm pv})^2}$.

\begin{figure}
\begin{center}
\includegraphics[width=0.53\textwidth]{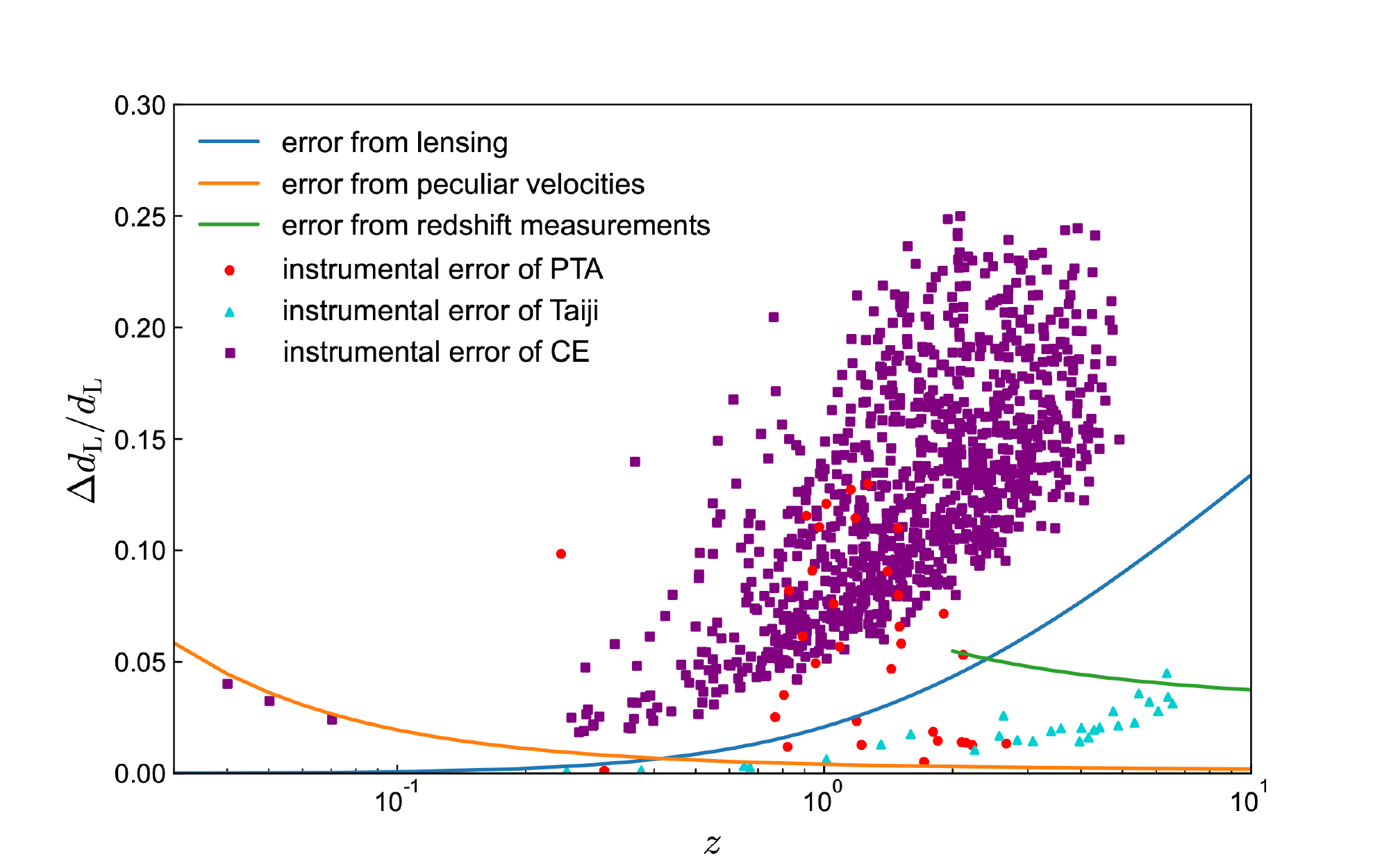}
\caption{The relative errors of luminosity distances from lensing, peculiar velocities, redshift measurements (only for GW standard sirens at $z>2$ from Taiji), and instrumental errors of PTA, Taiji, and CE.}\label{Fig1}
\end{center}
\end{figure}

\begin{figure}
\begin{center}
\includegraphics[width=0.45\textwidth]{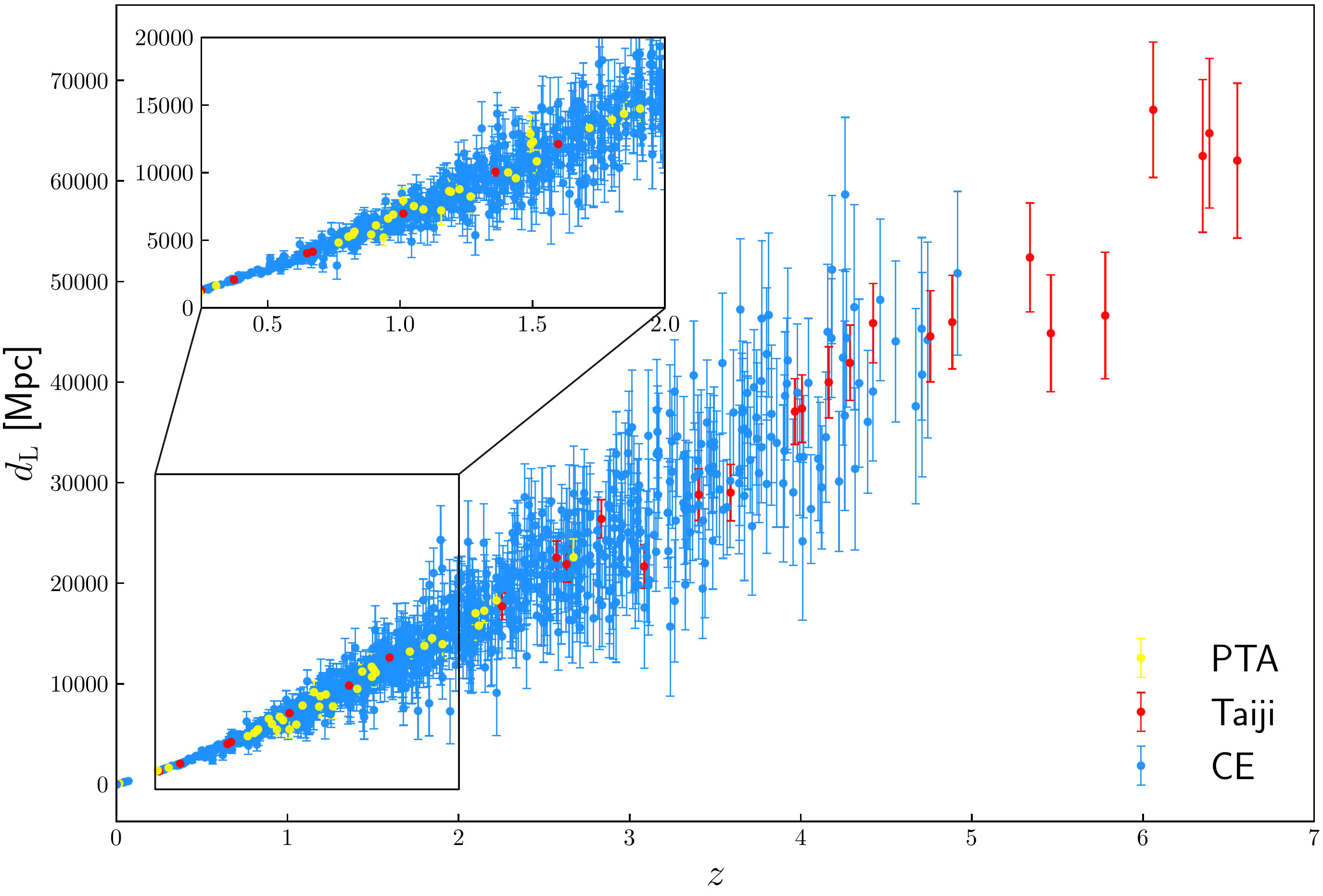}
\caption{The simulated GW standard siren data points observed by PTA, Taiji, and CE. The yellow data points represent the 35 standard sirens from the 10-year observation of PTA, the red data points represent the 28 standard sirens from the 5-year observation of Taiji, and the blue data points represent the 1000 standard sirens from the 10-year observation of CE.}\label{Fig2}
\end{center}
\end{figure}

Recent forecasts show that the 3G ground-based GW detectors would detect $\mathcal{O}(10^5)$ BNS mergers per year, but only about $0.1\%$ of them have the detectable EM counterparts \cite{Yu:2021nvx}. Chen \emph{et al.} recently showed that 910 GW standard sirens could be detected based on the 10-year observation of CE and Swift++ \cite{Chen:2020zoq}. Therefore, in the forecast in the present work, we simulate 1000 GW standard sirens generated by BNS mergers based on the 10-year observation of CE.

For each simulated standard siren event, the masses of NSs ($m_{1}$, $m_{2}$) are randomly chosen in the ranges of [1, 2] $M_{\odot}$. Without loss of generality, the merger time is chosen to $t_{\rm c}=0$ in our analysis. Here we note that the inclination angle should be randomly chosen in the range of $\cos\iota$ $\in$ $[-1, 1]$ when simulating isotropic GW sources. However, in this work, we assume that the redshifts of the GW events are determined by detecting SGRBs. Since SGRBs are strongly beamed, the detectable inclination angle is about $\iota\leq$ $20^{\circ}$ \cite{VERITAS:2021imb,Chen:2020zoq,Hirata:2010ba,Speri:2020hwc,Kocsis:2005vv,Rezzolla:2011da}. Therefore, in the present work, $\iota$ is randomly chosen in the range of $\iota\in [0,20^\circ]$. The simulation of other angles is the same as described in the simulation of MBHB.
{Based on the above analysis, we simulate 1000 bright sirens for the 10-year observation of CE, as shown in Fig.~\ref{Fig2}.}

{In Fig.~\ref{Fig1}, we present the relative errors of luminosity distances due to lensing, peculiar velocities, redshift measurements (only for GW standard sirens at $z>2$ from Taiji), and instrumental error of PTA, Taiji, and CE. For PTA, both the instrumental error and lensing error dominate the error of $d_{\rm L}$. On the other hand, for Taiji, the error of $d_{\rm L}$ is dominated by lensing due to the high SNRs of GW events from Taiji. For CE, the instrumental error dominates the error of $d_{\rm L}$. We combine the errors for every standard siren and construct the Hubble diagram using the future multi-band standard siren observations.} In Fig.~\ref{Fig2}, we show the simulated standard siren data from PTA, Taiji, and CE. We can see that the number of standard sirens from CE is the largest, followed by PTA and Taiji. However, due to the fact that SNRs of GW events observed by CE are lower than those of PTA and Taiji, CE has the largest errors of luminosity distances at the same redshifts. The luminosity distance errors of PTA and Taiji are almost the same at similar redshifts.

{We adopt the Markov Chain Monte Carlo method \cite{Lewis:2002ah} to maximize the likelihood $\mathcal{L}\propto\exp(-\chi^2/2)$ and infer the posterior probability distributions. The $\chi^2$ function is defined as
\begin{equation}
\chi^2=\sum_{i=1}^{N}\left(\frac{d_{{\rm L},i}^{\rm obs}-d_{{\rm L},i}^{\rm th}}{\sigma_{d_{{\rm L},i}}}\right)^2,
\end{equation}
where $N$ is the number of standard siren data points.}

\begin{table*}
\renewcommand\arraystretch{1.5}
\caption{The absolute errors (1$\sigma$) and the relative errors of the cosmological parameters in the $\Lambda$CDM, $w$CDM, and $w_{0}w_{a}$CDM models using the CMB, PTA, Taiji, CE, PTA+Taiji+CE, and CMB+PTA+Taiji+CE data. Here $H_0$ is in units of km s$^{-1}$ Mpc$^{-1}$.}\label{Tab1}
\centering
\normalsize
\setlength{\tabcolsep}{3mm}{
\begin{tabular}
{p{1.2cm}<{\centering} p{1.2cm}<{\centering} p{1.2cm}<{\centering} p{1.2cm}<{\centering} p{1.2cm}<{\centering} p{1.2cm}<{\centering} p{2.5cm}<{\centering} p{3.4cm}<{\centering}}
\hline\hline
%\bottomrule[1pt]
Model & Error & CMB & PTA  & Taiji & CE  & PTA+Taiji+CE & CMB+PTA+Taiji+CE \\
\hline
 \multirow{4}{*}{$\Lambda$CDM}&  $\sigma(\Omega_m)$& 0.009 & 0.020 & 0.024 & 0.012 & 0.008 &0.003 \\
& $\sigma(H_0) $   & 0.61  & 0.49 & 0.70 & 0.45 & 0.29 &0.20\\
%\hline
 & $\varepsilon(\Omega_m)$ &$2.7\%$ & $6.3\%$ & $11.7\%$ & $3.8\%$ & $2.8\%$  & $1.0\%$\\
  & $\varepsilon(H_0)$  &$0.9\%$ & $0.7\%$ & $1.6\%$ & $0.7\%$ & $0.5\%$ & $0.3\%$ \\
\hline
\multirow{6}{*}{$w$CDM}&  $\sigma(\Omega_m)$ & 0.058 & 0.038 & 0.033 & 0.017  & 0.015 & 0.003\\
  & $\sigma(H_0) $  & 6.30   & 2.25 & 1.75 & 0.95 & 0.69 & 0.37\\
& $\sigma(w) $  & 0.210 & 0.395 & 0.235 & 0.120 & 0.101 & 0.028 \\
%\hline
 & $\varepsilon(\Omega_m)$ & $18.5\%$ & $11.8\%$ & $10.2\%$ & $5.4\%$ & $4.7\%$ & $1.0\%$ \\
  & $\varepsilon(H_0)$  & $9.1\%$    & $3.3\%$ & $2.6\%$ & $1.4\%$ & $1.0\%$ & $0.6\%$ \\
 & $\varepsilon(w)$   &$20.2\%$    & $31.9\%$ & $22.0\%$ & $11.8\%$ & $10.0\%$ & $2.8\%$ \\
\hline
 \multirow{7}{*}{$w_{0}w_{a}$CDM} &  $\sigma(\Omega_m)$ & 0.066 & 0.069 & 0.059 & 0.048 & 0.047 & 0.009\\
  & $\sigma(H_0) $  & 7.25   & 4.30 & 3.10 & 1.40 & 1.35 & 0.85 \\
 & $\sigma(w_{0}) $  & 0.605  & 0.750 & 0.530 & 0.220 & 0.195 & 0.110 \\
%\hline
& $\sigma(w_{a})$ & 2.50 & 2.90 & 2.80 & 1.31 & 1.22 & 0.32\\
 & $\varepsilon(\Omega_m)$ &$20.6\%$ & $19.9\%$ & $16.9\%$ & $14.8\%$ & $14.8\%$ & $2.7\%$\\
  & $\varepsilon(H_0)$   & $10.6\%$   & $6.4\%$ & $4.7\%$ & $2.1\%$ & $2.0\%$ & $1.3\%$ \\
 & $\varepsilon(w_{0})$   & $112.0\%$   & $76.5\%$ & $74.6\%$ & $24.2\%$ & $21.0\%$  &$10.9\%$ \\
\hline\hline
%\bottomrule[1pt]
\end{tabular}}
\end{table*}

\section{Results and discussion}\label{results}

\begin{figure*}
\includegraphics[width=0.4\textwidth]{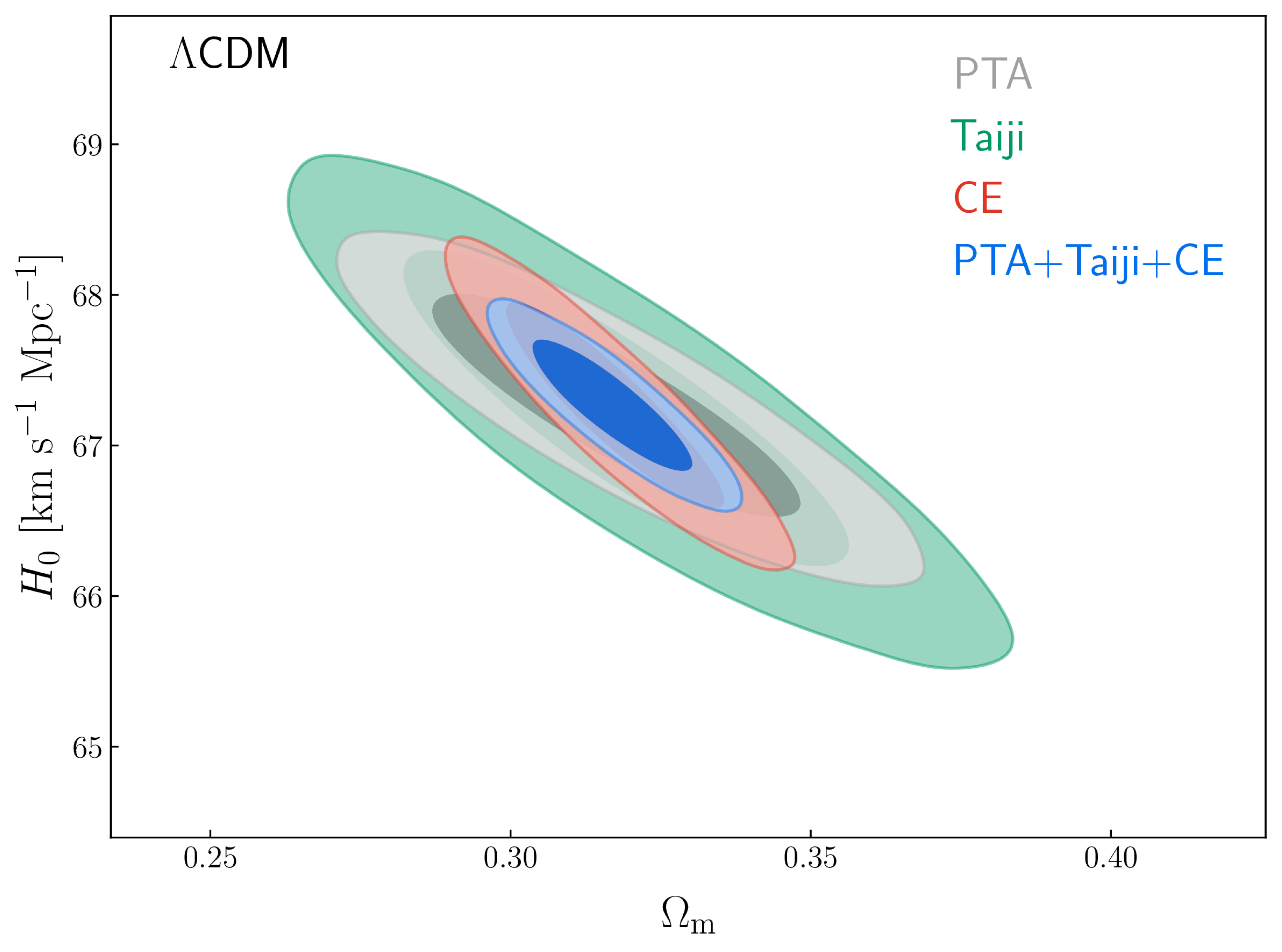}%scale=0.55
\includegraphics[width=0.4\textwidth]{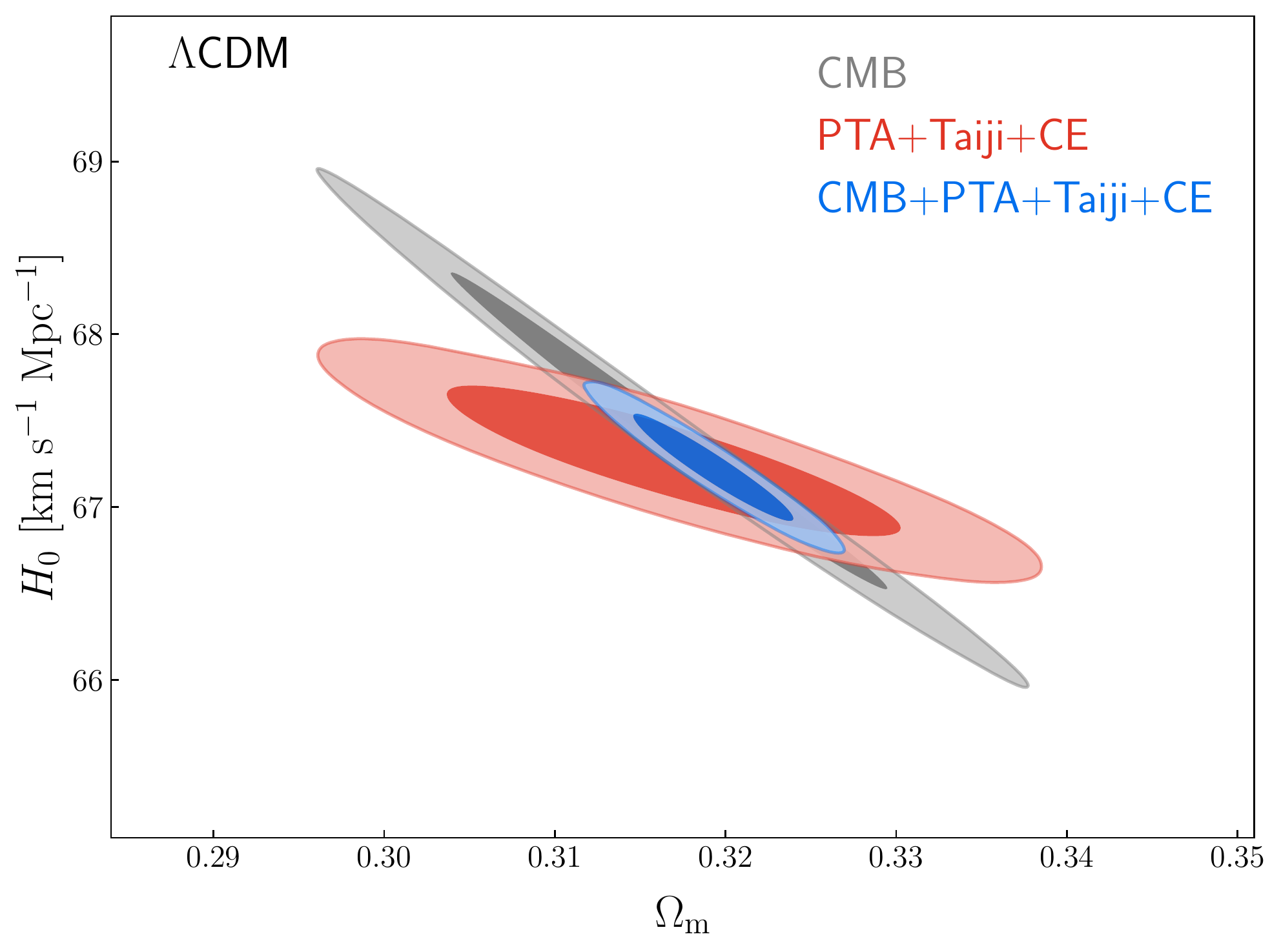}%scale=0.55
\caption{Constraints on the $\Lambda$CDM model. Left panel: Two-dimensional marginalized contours ($68.3\%$ and $95.4\%$ confidence level) in the $\Omega_{\rm m}$--$H_0$ plane by using the PTA, Taiji, CE, and PTA+Taiji+CE data. Right panel: Two-dimensional marginalized contours ($68.3\%$ and $95.4\%$ confidence level) in the $\Omega_{\rm m}$--$H_0$ plane by using the CMB, PTA+Taiji+CE, and CMB+PTA+Taiji+CE data.}\label{Fig3}
\end{figure*}

\begin{figure*}
\includegraphics[width=0.4\textwidth]{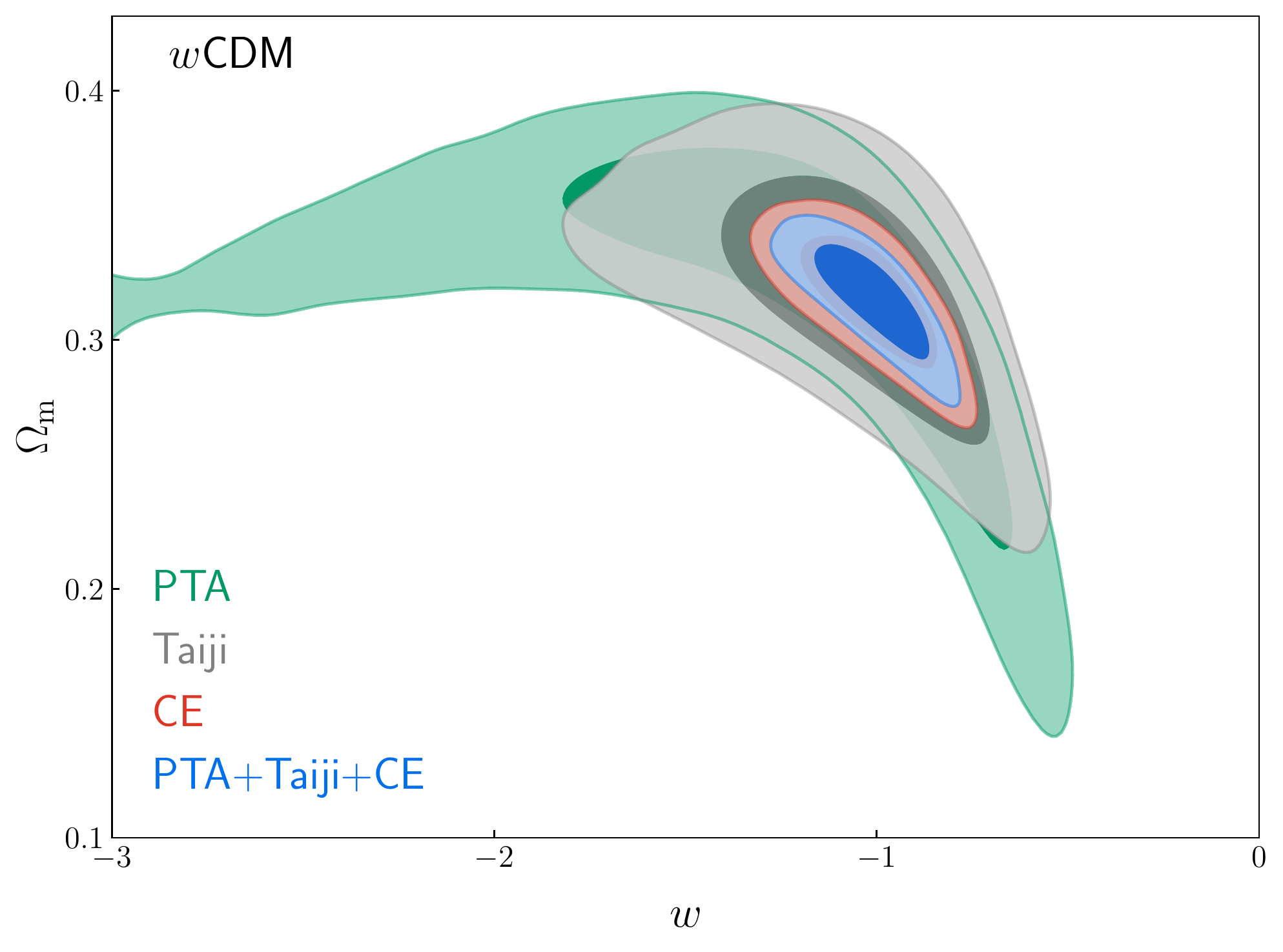}%scale=0.55
\includegraphics[width=0.4\textwidth]{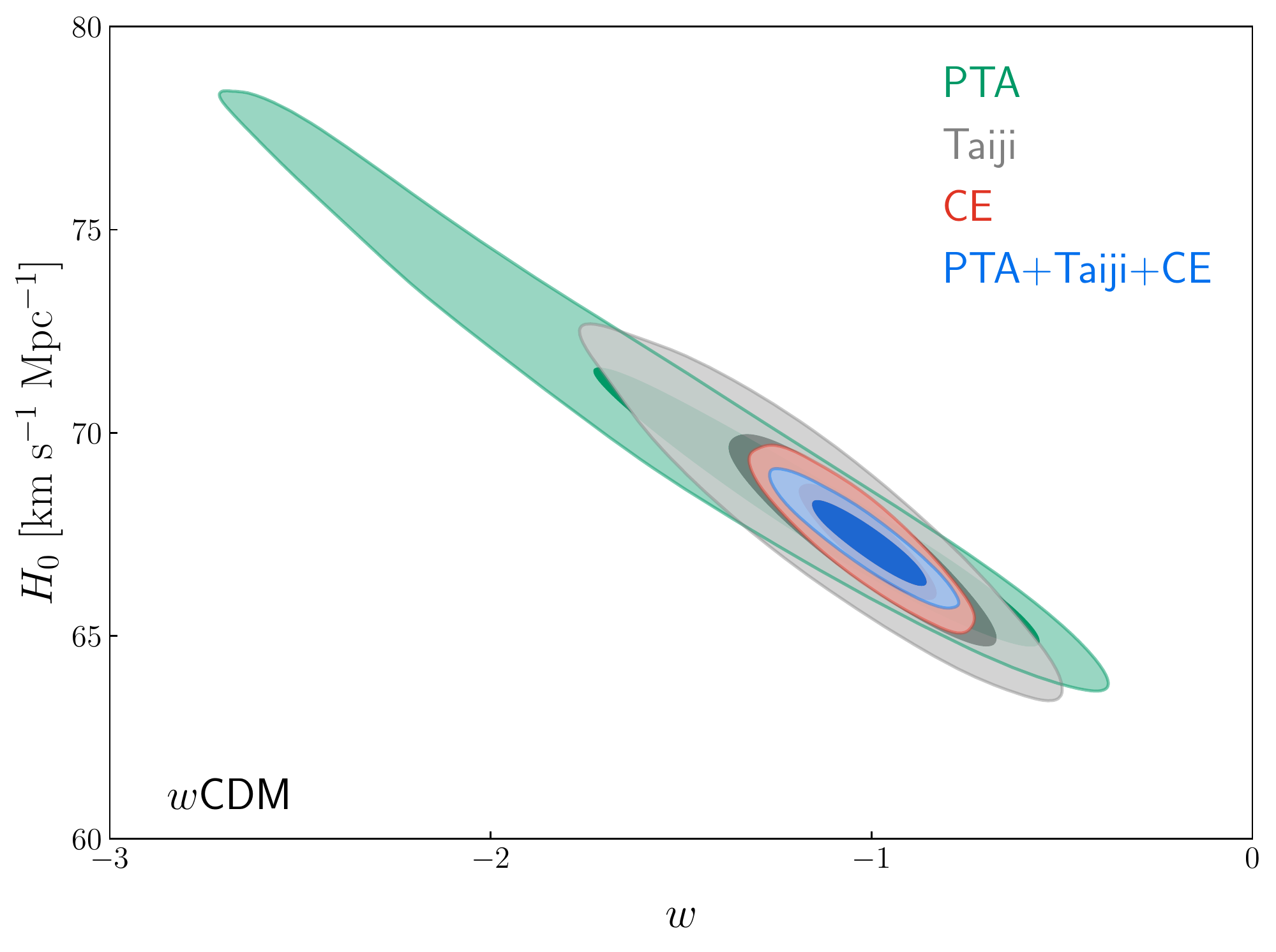}%scale=0.55
\caption{Constraints on the $w$CDM model. Here we show the two-dimensional marginalized contours ($68.3\%$ and $95.4\%$ confidence level) in the $w$--$\Omega_{\rm m}$ (left panel) and $w$--$H_0$ (right panel) planes using the PTA, Taiji, CE, and PTA+Taiji+CE data.}\label{Fig4}
\end{figure*}

\begin{figure*}
\includegraphics[width=0.4\textwidth]{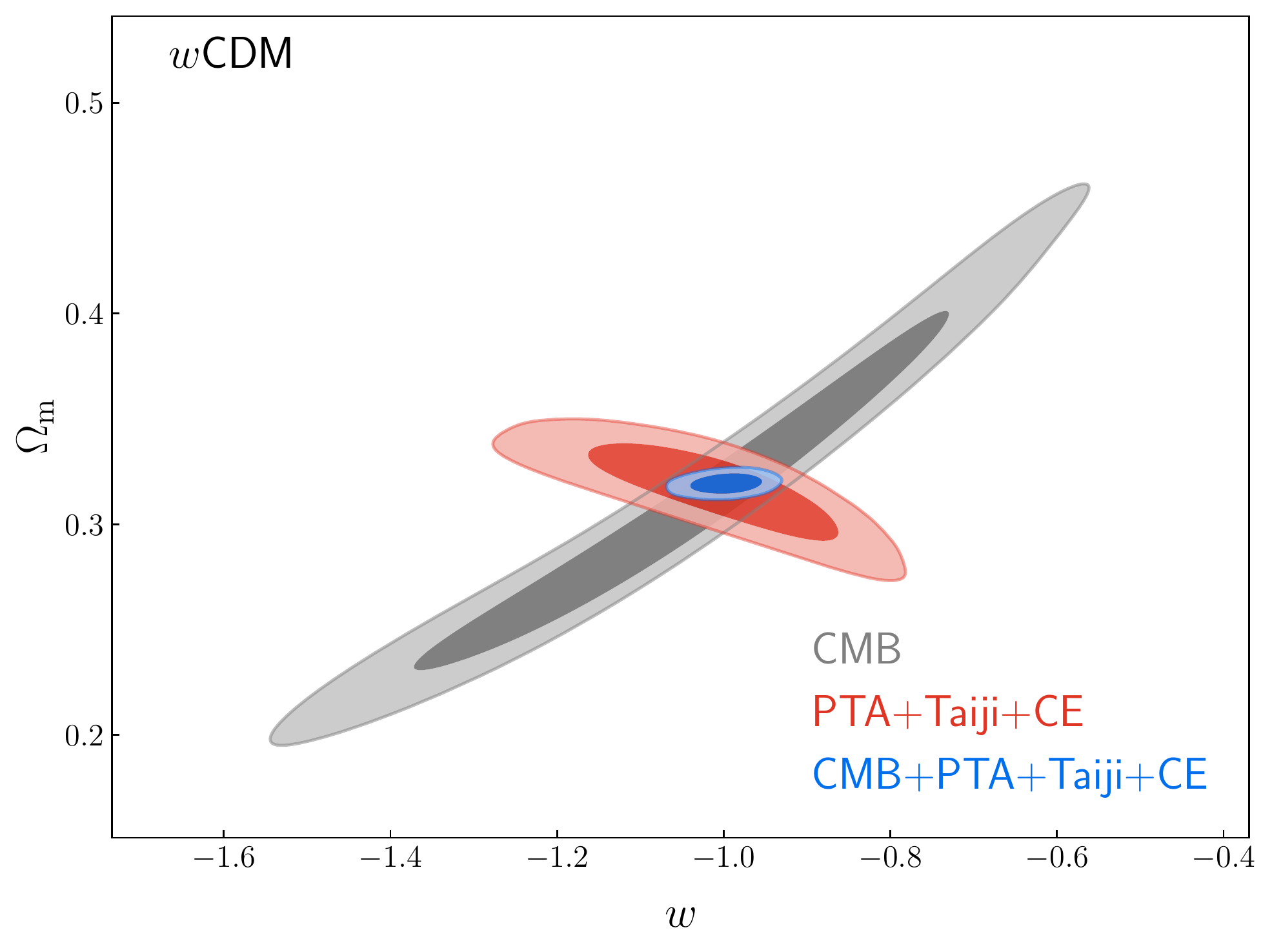}%scale=0.55
\includegraphics[width=0.4\textwidth]{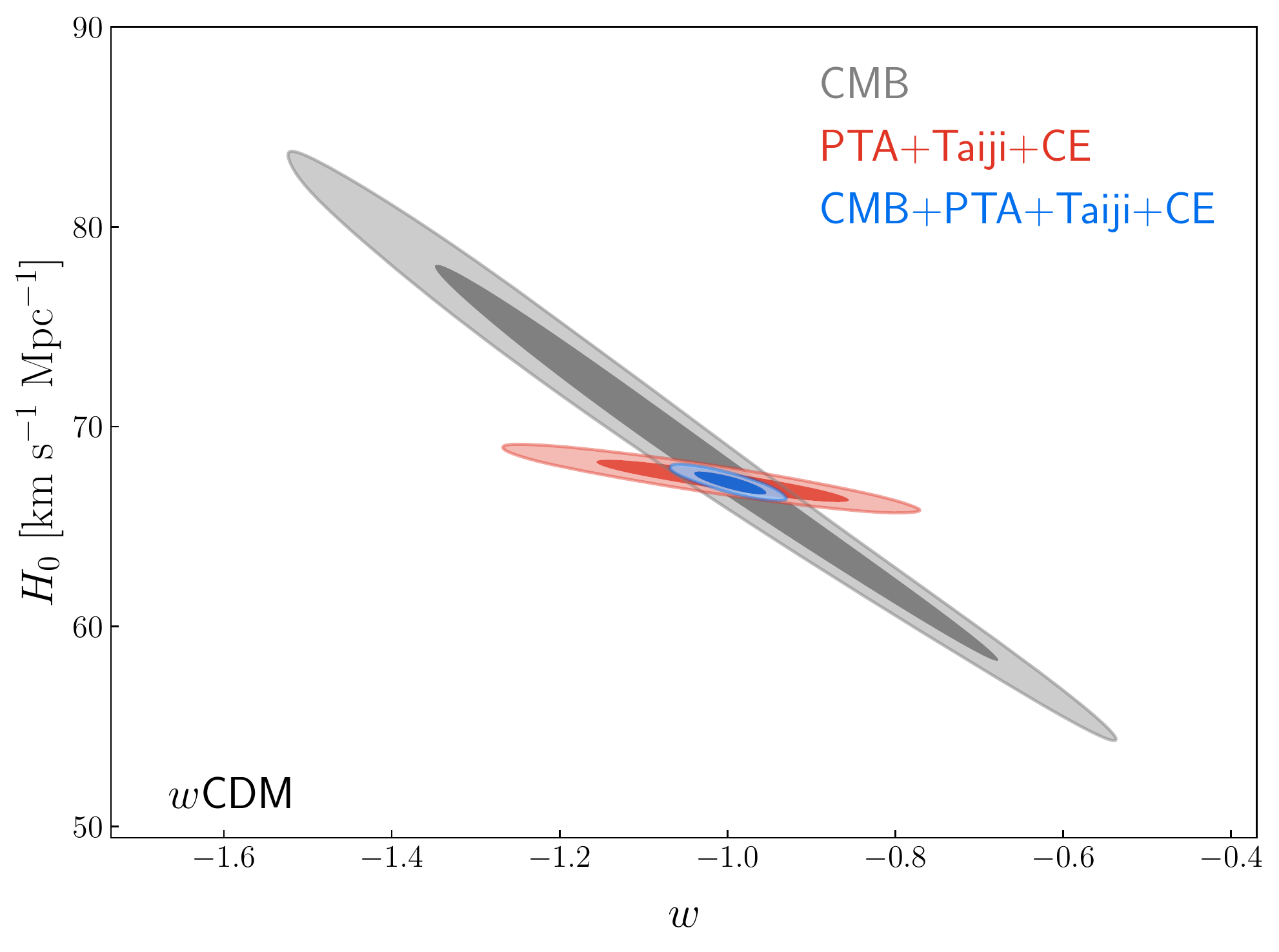}%scale=0.55
\caption{Constraints on the $w$CDM model. Here we show the two-dimensional marginalized contours ($68.3\%$ and $95.4\%$ confidence level) in the $w$--$\Omega_{m}$ (left panel) and $w$--$H_{0}$ (right panel) planes using the CMB, PTA+Taiji+CE, and CMB+PTA+Taiji+CE data.}\label{Fig5}
\end{figure*}

\begin{figure*}[!htbp]
\includegraphics[width=0.4\textwidth]{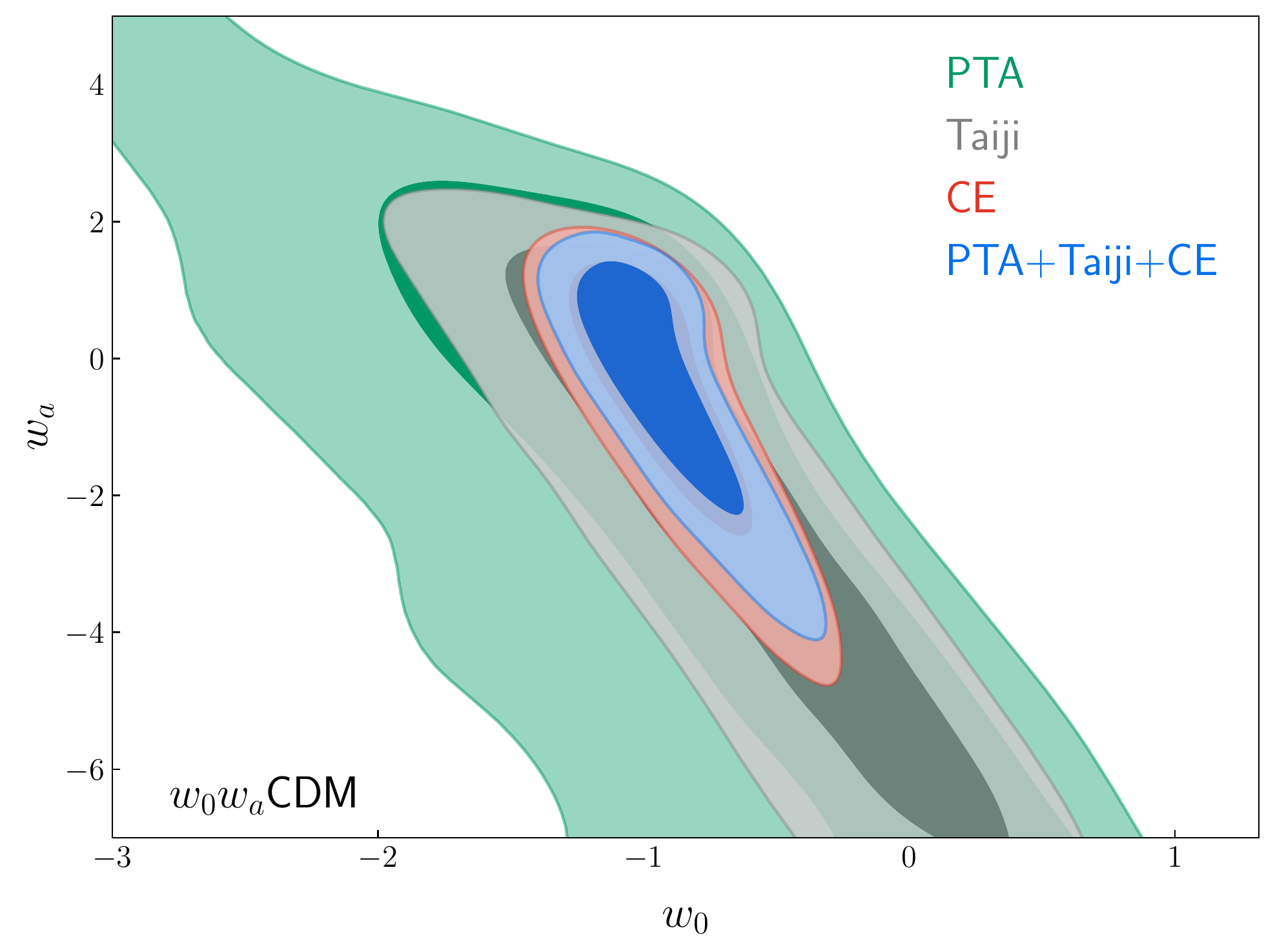}%scale=0.55
\includegraphics[width=0.4\textwidth]{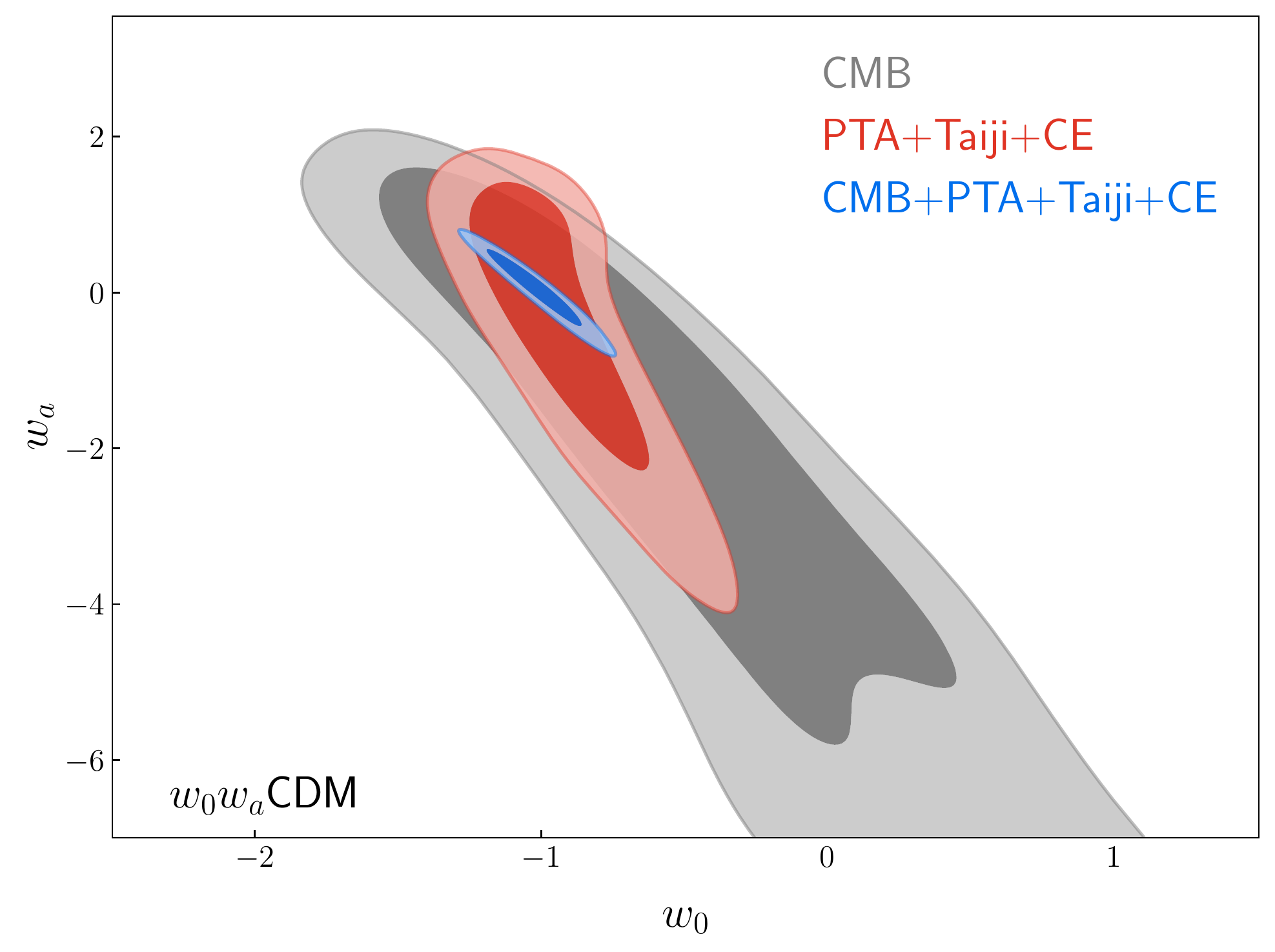}%scale=0.55
\caption{Constraints on the $w_0w_a$CDM model. Left panel: Two-dimensional marginalized contours ($68.3\%$ and $95.4\%$ confidence level) in the $w_0$--$w_a$ plane by using the PTA, Taiji, CE, and PTA+Taiji+CE data. Right panel: Two-dimensional marginalized contours ($68.3\%$ and $95.4\%$ confidence level) in the $w_0$--$w_a$ plane by using the CMB, PTA+Taiji+CE, and CMB+PTA+Taiji+CE data.}\label{Fig6}
\end{figure*}

In this section, we report the constraint results. We use the simulated GW standard siren data from PTA, Taiji, and CE to constrain the $\Lambda$CDM [$w(z)=-1$], $w$CDM [$w(z)=\rm constant$], and $w_0w_a$CDM [$w(z)=w_0+w_a z/(1+z)$] models by performing the Markov-chain Monte Carlo analysis \cite{Lewis:2002ah}. For the CMB data, we employ the ``$Planck$ distance priors'' from the $Planck$ 2018 observation \cite{Chen:2018dbv,Planck:2018vyg}. The $1\sigma$ and $2\sigma$ posterior distribution contours for the cosmological parameters of interest are shown in Figs.~\ref{Fig3}--\ref{Fig6} and the $1\sigma$ errors for the marginalized parameter constraints are summarized in Table~\ref{Tab1}. We use $\sigma(\xi)$ and $\varepsilon(\xi)$ to represent the 1$\sigma$ absolute and relative errors of the parameter $\xi$, with $\varepsilon(\xi)$ defined as $\varepsilon(\xi)=\sigma(\xi)/\xi$.

We first focus on the constraint results for the $\Lambda$CDM model. In the left panel of Fig.~\ref{Fig3}, we show the constraint results in the $\Omega_{\rm m}$--$H_0$ plane by using the simulated PTA, Taiji, CE, and PTA+Taiji+CE data. As can be seen, CE contributes the most to the PTA+Taiji+CE results, followed by PTA and Taiji. This is because the number of simulated standard sirens from CE is much more than those of PTA and Taiji. Although the measurement errors of $d_{\rm L}$ for CE are large, the constraints on cosmological parameters are reduced statistically. Compared with Taiji, PTA has more lower-redshift data points ($z<2$), so PTA can better constrain the Hubble constant. Moreover, due to the different redshift intervals of the simulated PTA, Taiji, and CE data, their parameter degeneracy orientations are slightly different, so their combination could break cosmological parameter degeneracies.
The combination of PTA, Taiji, and CE gives $\sigma(\Omega_{\rm m})=0.008$ and $\sigma(H_0)=0.29$ km s$^{-1}$ Mpc$^{-1}$, which are $33.3\%$ [$(0.012-0.008)/0.012$] and $35.6\%$ [$(0.45-0.29)/0.45$] better than those of CE. In the right panel of Fig.~\ref{Fig3}, we can see that the contours of CMB and PTA+Taiji+CE show different orientations and thus the combination of them could break cosmological parameter degeneracies. The prime cause is that GW could measure $H_0$ better, so it could lead to a different degeneracy direction compared with CMB. The combination of CMB and PTA+Taiji+CE gives $\sigma(\Omega_{\rm m})=0.003$ and $\sigma(H_0)=0.20$ km s$^{-1}$ Mpc$^{-1}$, and the constraint precisions of $\Omega_{\rm m}$ and $H_0$ are $1.0\%$ and $0.3\%$, meeting the standard of precision cosmology. In general, the joint PTA+Taiji+CE data could tightly constrain the Hubble constant, and if combined with CMB, the measurement precisions of cosmological parameters could be greatly improved (better than or at least equal to $1\%$), due to the parameter degeneracies being broken.

In Figs.~\ref{Fig4} and \ref{Fig5}, we show the constraint results in the $w$--$\Omega_{\rm m}$ and $w$--$H_0$ planes for the $w$CDM model. As can be seen from Fig.~\ref{Fig4}, CE also contributes the most to the PTA+Taiji+CE data. However, the ability of Taiji to constrain cosmological parameters in the $w$CDM model is better than that of PTA. This is because Taiji has more high-redshift standard sirens than PTA, so Taiji can better constrain EoS parameter of dark energy $w$. Meanwhile, the combination of PTA, Taiji, and CE gives $\sigma(w)=0.101$, which is $15.8\%$ better than the constraint result by CE. In Fig.~\ref{Fig5}, we see that the parameter degeneracy orientations of CMB and PTA+Taiji+CE are almost orthogonal and thus the combination of them could not only break cosmological parameter degeneracies but also tremendously improve the cosmological parameter constraints. The addition of the PTA+Taiji+CE data could reduce the $1\sigma$ absolute error of $w$ by $86.7\%$, compared with CMB. Moreover, the combination of CMB and PTA+Taiji+CE gives $\sigma(w)=0.028$, which is comparable with the latest constraint result by the CMB+BAO+SN data \cite{Brout:2022vxf}.

In Fig.~\ref{Fig6}, we show the case for the $w_0w_a$CDM model in the $w_0$--$w_a$ plane. As can be seen from the left panel of Fig.~\ref{Fig6}, the constraint results are the same as those in the $w$CDM model, i.e., CE contributes the most, followed by Taiji and PTA. The joint PTA+Taiji+CE data could give $\sigma(w_0)=0.195$ and $\sigma(w_0)=1.22$, which are both better than the constraint results by the CMB data. Furthermore, in the right panel of Fig.~\ref{Fig6}, we see that the combination of CMB and PTA+Taiji+CE could also break the parameter degeneracies and thus significantly improve the cosmological parameter constraints. The joint CMB+PTA+Taiji+CE data give $\sigma(w_0)=0.110$ and $\sigma(w_a)=0.32$, which are $81.8\%$ [$(0.605-0.110)/0.605$] and $87.2\%$ [$(2.50-0.32)/2.50$] better than the results of CMB.

Our results show that the joint future multi-band GW standard siren observations would play a crucial role in cosmological parameter estimation. CE contributes the most to the PTA+Taiji+CE results since the number of standard sirens detected by CE is much more than those of PTA and Taiji. PTA has more lower-redshift ($z<2$) data. Taiji has more higher-redshift data. Hence, PTA offers better constraints in the $\Lambda$CDM model, while in the dynamical dark energy models, Taiji offers better constraints. The joint multi-band GW standard siren data show great potential in constraining the $\Lambda$CDM model. Moreover, the parameter degeneracy orientations of them are slightly different and thus the combination of them could break the cosmological parameter degeneracies. However, the joint constraints perform not well in the $w$CDM and $w_0w_a$CDM models. Fortunately, the joint PTA+Taiji+CE data have different parameter degeneracy orientations from CMB, so the combination of them could effectively break the parameter degeneracies and greatly improve constraint precisions of cosmological parameters. It can be concluded that the future multi-band GW observations are worth expecting in precisely measuring cosmological parameters and helping solve important cosmological problems. {Here we emphasize that the successful application of the standard siren method heavily depends on the accuracy of the GW data \cite{LIGOScientific:2017aaj,LIGOScientific:2019hgc}. In particular, the systematic errors caused by calibrations in the data processing should be carefully avoided. In this work, the systematic errors from calibrations in the standard siren data are not considered (for the impact of calibration uncertainties on the Hubble constant measurements, see e.g. Ref.~\cite{Huang:2022rdg}).}

\section{Conclusion}\label{Conclusion}
In this work, we explore the potential of the joint constraints on cosmological parameters using future multi-band GW standard siren observations. We simulated the multi-band standard siren data based on the 10-year observation of CE, the 5-year observation of Taiji, and the 10-year observation of the SKA-era PTA, and used mock data to constrain three typical cosmological models, i.e., the $\Lambda$CDM, $w$CDM, and $w_0w_a$CDM models.

We find that the joint PTA+Taiji+CE data could give tight constraints on the Hubble constant, with the constraint precision being $0.5\%$ in the $\Lambda$CDM model. However, the joint data perform not well in constraining EoS parameters of dark energy. Fortunately, CMB and PTA+Taiji+CE show different parameter degeneracy orientations, and thus the combination of them could effectively break the parameter degeneracies and improve constraint precisions of cosmological parameters. In the $\Lambda$CDM model, the constraint precisions of $\Omega_{\rm m}$ and $H_0$ using the CMB+PTA+Taiji+CE data are better than or at least equal to $1\%$. While in the $w$CDM model, CMB+PTA+Taiji+CE offers $\sigma(w)=0.028$, which is comparable with the latest constraint result by the CMB+BAO+SN data. Compared with CMB, the combination of CMB and PTA+Taiji+CE could improve the constraint on $w$ by $86.7\%$. In the $w_0w_a$CDM model, the CMB+PTA+Taiji+CE data offer $\sigma(w_0)=0.110$ and $\sigma(w_a)=0.32$, which are $81.8\%$ and $87.2\%$ better than the results by CMB.

Hence, we can conclude that: (i) the joint future multi-band GW standard sirens could precisely measure the Hubble constant, but are not good at measuring dark energy; (ii) the joint PTA+Taiji+CE data could effectively break the cosmological parameter degeneracies generated by the CMB data, especially in the dynamical dark energy models. It is worth expecting to use the future multi-band GW observations to probe the nature of dark energy and measure the Hubble constant.

\section*{Acknowledgements}
We thank Yong Yuan, Tao Han, and Peng-Ju Wu for helpful discussions.
This work was supported by the National SKA Program of China (Grants Nos. 2022SKA0110200 and 2022SKA0110203) and the National Natural Science Foundation of China (Grants Nos. 11975072, 11875102, and 11835009).

\bibliography{gwmultiband}

\end{document}